\documentclass[aip,apl,reprint,superscriptaddress,citeautoscript,floatfix]{revtex4-1}
\usepackage{amsmath}
\usepackage{graphicx}
\usepackage{epstopdf}
\usepackage{hyperref}
\usepackage{color,soul}

\let\oldhat\hat

\renewcommand{\hat}[1]{\oldhat{\mathbf{#1}}}
\newcommand{\upperRomannumeral}[1]{\uppercase\expandafter{\romannumeral#1}}

\begin{document}

%\preprint{}

\title{Imaging collective behavior in an rf-SQUID metamaterial tuned by DC and RF magnetic fields}

\author{Alexander P. Zhuravel}
\affiliation{B. Verkin Institute for Low Temperature Physics and Engineering, Kharkov  61103, Ukraine}

\author{Seokjin Bae}
\affiliation{Center for Nanophysics and Advanced Materials, Department of Physics, University of Maryland, College Park, MD 20742, USA}

\author{Alexander V. Lukashenko}
\affiliation{Physikalisches Institut, Karlsruhe Institute of Technology, 76131 Karlsruhe, Germany}

\author{Alexander S. Averkin}
\affiliation{Russian Quantum Center, National University of Science and Technology MISIS, Moscow 119049, Russia}

\author{Alexey V. Ustinov}
\affiliation{Physikalisches Institut, Karlsruhe Institute of Technology, 76131 Karlsruhe, Germany}
\affiliation{Russian Quantum Center, National University of Science and Technology MISIS, Moscow 119049, Russia}

\author{Steven M. Anlage}
\affiliation{Center for Nanophysics and Advanced Materials, Department of Physics, University of Maryland, College Park, MD 20742, USA}

\date{\today}

\begin{abstract}
We examine the collective behavior of two-dimensional nonlinear superconducting metamaterials using a non-contact spatially resolved imaging technique. The metamaterial is made up of sub-wavelength nonlinear microwave oscillators in a strongly coupled 27$\mathtt{\times}$27 planar array of radio-frequency Superconducting QUantum Interference Devices (rf-SQUIDs). By using low-temperature laser scanning microscopy we image microwave currents in the driven SQUIDs while in non-radiating dark modes and identify the clustering and uniformity of like-oscillating meta-atoms. We follow the rearrangement of coherent patterns due to meta-atom resonant frequency tuning as a function of external dc and rf magnetic flux bias. We find that the rf current distribution across the SQUID array at zero dc flux and small rf flux reveals a low degree of coherence. By contrast, the spatial coherence improves dramatically upon increasing of rf flux amplitude, in agreement with simulation. 
\end{abstract}

%\keywords{}

\maketitle
%\indent

\par Planar arrays of deep sub-wavelength dimension superconducting (SC) resonators have recently gained increasing attention due to their potential use as nonlinear metamaterials, i.e., engineered media whose electromagnetic response can differ dramatically from natural materials. Arrays of coupled SC resonators can be used for controllable routing and manipulating electromagnetic wave propagation in the range from radio-frequency (rf)\cite{Kurter2010APL,Anlage2011JOpt} through the THz\cite{Zheludev2012NatMat,Scalari2014APL} domain. Demonstrated effects include electromagnetically\cite{Kurter2012APL} and self-induced broadband transparency,\cite{Daimeng2015PRX} negative magnetic permeability,\cite{Butz2013OptExp} polarization rotation,\cite{Caputo2015PRB} Fano resonance,\cite{Lukyanchuk2010NatMat} and multi-stable states.\cite{Jung2014NatComm} Extensive progress on the development and applications of SC metamaterials has been achieved.\cite{Ricci2005APL, CCDu2006PRB, Rakhmanov2008PRB, Anlage2011JOpt, Kurter2011IEEE, Savinov2012PRL, Jung2014SST, Ustinov2015IEEE, Averkin2017IEEE, Lazarides2018PRE,Lazarides2018PhysRep} It has been shown that such SC structures have significant advantages over their normal-metal counterparts allowing reduced losses by several orders of magnitude, shrinking the size of artificial meta-atoms, and achieving tunable frequency of operation by means of macroscopic quantum phenomena.\cite{Jung2014SST,Lazarides2018PhysRep}

\par Traditional SC metamaterials are composed of ultra-compact, self-resonating spirals and split-ring resonators (SRR) of different design.\cite{Kurter2010APL, Savinov2012PRL, Jung2014SST, Maleeva2015JAP, Averkin2017IEEE} Potentially, these structures could be used for switching,\cite{Zheludev2010Science, Kurter2012APL,Jung2014NatComm} as well as for tuning, by virtue of the dynamic range of well controllable manipulation of their resonances with external stimuli.\cite{RicciIEEE2007,Lapine2014RMP} 
\cite{}
\par There has been great interest in using radio-frequency Superconducting QUantum Interference Devices: rf-SQUIDs as meta-atoms.\cite{Ricci2005APL, CCDu2006PRB, Lazarides2007APL, Zheludev2010Science, Maimistov2010OptCommun, Anlage2011JOpt, Butz2013OptExp, Jung2014SST, Lazarides2018PRE,  Lazarides2018PhysRep} Contrary to the SRR and SC spiral, the rf-SQUID incorporates an extremely tunable nonlinear inductor, arising from the Josephson effect, when a Josephson junction is incorporated into a SC loop. The loop geometry adds the macroscopic quantum property of flux quantization, which in turn allows one to conveniently control the Josephson inductance through the magnetic flux applied to the loop.\cite{Trepanier2013PRX}  The benefits of the rf-SQUID system were originally presented theoretically in the context of an artificial atom with discrete energy levels\cite{CCDu2006PRB} and extremely strong nonlinearity.\cite{Lazarides2007APL, Caputo2012PRB} This system can exhibit negative and/or oscillating effective magnetic permeability which can be wide-band tuned together with the resonance.  

\par Recent experiments on rf-SQUID-based meta-atoms have lived up to most if not all of the theoretical expectations.\cite{Ustinov2015IEEE,Lazarides2018PhysRep} It has been shown that the rf-SQUID meta-atoms have a rich nonlinear behavior arising from the nonlinearity of the Josephson junction.\cite{Jung2014NatComm,Daimeng2016PRB} In the case of small rf drive, the tuning capability of the single rf-SQUID is an $n\Phi_0$-periodic function ($n$ is an integer) of the flux quantum $\Phi_0$=$h/2e$, while dc magnetic flux ($\Phi_{dc}$) dependent variations of the resonant frequency can reach up to 80 THz/Gauss.\cite{Trepanier2013PRX, Jung2013APL, Butz2013SST, Jung2013APL} Furthermore, the rf-SQUID demonstrates undistorted, high-Q resonances under tuning by $\Phi_{dc}$.\cite{Vidiborskiy2013APL, Trepanier2013PRX, Jung2013APL, Butz2013SST, Jung2013APL} However, degradation of dc flux tunability and hysteresis behavior has been documented at increased temperature $T$ and rf flux amplitude $\Phi_{rf}$.\cite{Trepanier2013PRX, Jung2014NatComm, Daimeng2015PRX} 

\par Since the advantages of classical rf-SQUID-based meta-materials are defined by a collective response of oscillating meta-atoms, their natural and/or forced frequency (and phase) synchronization accompanied with coherent tunability are of paramount importance. As with any nonlinear discrete structure, there is a competition between self-organization and destruction of the collective resonant behavior due to disorder\cite{Jenkins2018PRB} and nonlinearity. The degree of spatial-temporal coherence (i.e., frequency and phase synchronization of the meta-atoms)\cite{Papasimakis2009PRB} of rf-SQUID metamaterials was previously examined experimentally and theoretically in terms of the collective globally-averaged transmission properties.\cite{Trepanier2017PRE} However, many magneto-inductive modes of the metamaterial exist that have very weak coupling to a uniform microwave excitation.\cite{Trepanier2015Thesis} Examination of these dark modes,\cite{Jenkins2018PRB} and their nonlinear evolution requires a microscopic form of investigation.

\par The phenomenon of incoherence in large populations of interacting SQUIDs is the subject of intense theoretical and numerical research,\cite{Cawthorne1999PRB, Acebron2005RMP, Lazarides2013SST, Trepanier2013PRX, Lazarides2015PRB, Caputo2015PRB,Trepanier2017PRE} and in particular, there are clear predictions for the development of Chimera states in rf-SQUID arrays\cite{Lazarides2018PhysRep, Hizanidis2016PRE} (see Refs.\cite{Panaggio2015Nonlin, Lazarides2018PRE} for a review). Here, we present images of the microscopic states and site-dependent coherence of nonlinear and disordered arrays of strongly coupled rf-SQUIDs. We address several important questions: (1) Can coherence be induced by large-amplitude rf driving flux? (2) Does the structure of synchronously coupled rf-SQUIDs remain stable under dc flux tuning? (3) How is this structure spatially modified in the presence of disorder, (e.g., due to randomly distributed imperfections during fabrication) and dc flux gradient?

\begin{figure}
		\includegraphics[width=0.8\columnwidth]{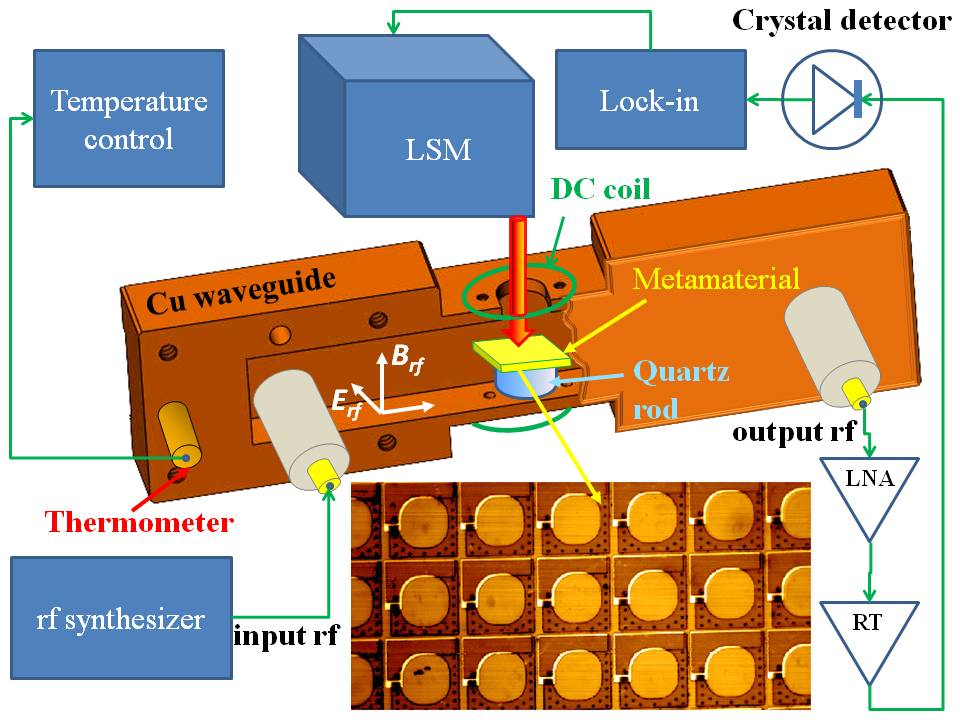}
		\caption{\label{fig:Fig1} Schematic representation of the LSM setup used for 2D visualization of microwave photo-response of the rf-SQUID metamaterial. LNA is the low-noise cryogenic amplifier while RT is a room temperature rf amplifier. The red line denotes the scanned laser beam while the green circles show the dc magnetic flux coils. Drawing is not to scale. The inset illustrates the LSM optical reflectivity image in a 500$\mathtt{\times}$250 $\mu$m$^2$  area of the array}
\end{figure}

\par The investigated SC metamaterial consists of 27$\mathtt{\times}$27 rf-SQUID meta-atoms equidistantly arranged in a square array with a lattice constant of 83 $\mu$m as shown in Fig.\ref{fig:Fig1} and Fig.S1 (Supp. Mat.). These rf-SQUIDs are formed on a Si substrate using trilayer Nb/AlO$_x$/Nb Josephson junction (JJ) technology. More details on array fabrication, microwave design, global experimental characterization and rf simulation of similar rf-SQUIDs arrays has been described in Ref.\cite{Yohannes2007IEEE, Trepanier2013PRX, Trepanier2015Thesis, Daimeng2016PRB} and Supp. Mat.

\par The spatially-resolved technique of Laser Scanning Microscopy (LSM) has been employed in the past to image rf current densities and local sources of nonlinearity through a controlled thermal perturbation of superconducting microwave circuits. Thermal perturbation of a dc nano-SQUID has been used before to image local dissipation,\cite{Halbertal2016Nature} and a preliminary study of rf-SQUID LSM has been published.\cite{Averkin2016IEEE} A simplified schematic of this setup is shown in Fig.\ref{fig:Fig1}. The rf-SQUID array is placed on a cooled quartz rod between the walls of a specially designed rectangular copper waveguide which has 14$\mathtt{\times}$7$\mathtt{\times}$80 mm$^3$ interior dimension (single propagating mode from 10.6 to 20 GHz). The rod fixes the array in a region of homogenous rf magnetic field in a way that the magnetic vector of the traveling TE$_{10}$ mode is predominantly perpendicular to the x-y plane of the SQUIDs. Note that the waveguide to coaxial cable transitions are impedance matched, and the single propagating mode provides the rf flux bias to the array. A circular opening in the waveguide wall provides LSM access to the rf-SQUIDs for 2D laser probing of their microwave properties, as well as visible-light reflectivity imaging, which allows for proper alignment of PR images. Two superconducting Helmholtz coils are glued to the waveguide to introduce a homogenous dc magnetic flux orthogonally threading the loops of the rf-SQUIDs. The entire construction is surrounded by mu-metal shielding to shield the rf-SQUIDs from environmental magnetic and electromagnetic influences. Additionally, its temperature is stabilized at $T_0 = 4.5$ K in a specialized optical cryostat with an accuracy of 1 mK, excluding any thermal drift of resonance frequency.

\par In this work, the rf-SQUIDs are stimulated by a low input rf power $P_{IN}=-60$ dBm, corresponding to rf flux amplitude $\Phi_{rf}\mathtt{\approx} 10^{-4}\Phi_0$.\cite{Trepanier2015Thesis} The sample is directly illuminated with a laser beam (wavelength 640 nm, laser power $P_L$=10 $\mu$W) that is focused into a 20 $\mu$m diameter spot, which is smaller than a single rf-SQUID but bigger than a Josephson junction. Its intensity is reduced within the LSM optics down to $3.2\mathtt{\times}10^2$ W/m$^2$ and TTL modulated at a frequency of $f_M\sim 1$ MHz while the absorbed radiation produces a periodic heating $\delta T<1$ mK underneath the laser probe (See Supp. Mat.). This laser-beam-induced perturbation is low enough not to change significantly the spatial distribution of resonating currents in the array. The probe scans point-by-point in a raster pattern over the examined area of the array. The maximum area of the LSM raster is 5$\mathtt{\times}$5 mm$^2$ while minimum step size between its discrete points is 1 $\mu$m.

\par Under these conditions the dominant effect of the laser beam perturbation is a change of the temperature dependent tunneling critical current $I_c$ of the probed JJ, resulting in the (oscillating at $f_M$) modification of its Josephson inductance $L_{JJ}(T,\Phi_{app}) = \frac{\Phi_0}{2\pi I_c \cos\delta}$, where $\delta$ is the gauge-invariant phase difference on the junction and $\Phi_{app}$=$\Phi_{dc}$+$\Phi_{rf}\sin(2\pi ft)$.  This causes, in turn, a periodic modulation of the resonant frequency $f_0$ of the rf-SQUID:
\begin{equation} \label{eq1}
f_0(T,\Phi_{app})=\frac{1}{2\pi\sqrt{\left(\frac{1}{L_{geo}}+\frac{1}{L_{JJ}(T,\Phi_{app})}\right)^{-1}C}},
\end{equation} 
that also oscillates under laser modulated heating.  Here $L_{geo}$ is the geometric inductance of the rf-SQUID loop and $C$ is the shunt capacitance of the JJ such that $f_{geo}$=$1/2\pi\sqrt{L_{geo}C}$ is the geometric resonance of the SQUID loop in the absence of a Josephson effect, and assuming that the electrodynamics of the SQUID is described by the resistively and capacitively shunted junction (RCSJ) model\cite{Likharev1986}. 

\par The laser beam induced differences between perturbed and unperturbed resonances causes a modulation of transmitted rf power PR$(\Phi_{app},f)$ $\mathtt{\sim}$ $\delta P_{OUT}(\Phi_{app},f)$ $\mathtt{\sim}$ $\delta||S_{21}(\Phi_{app},f)||^2$, where $S_{21}(f)$ is the spectral response of the forward transmission coefficient through the waveguide at frequency $f$. This signal is the LSM photo-response (PR) that is amplified by 67 dB, converted with a crystal diode to voltage $\delta V(t)$ signal and, after demodulation by a phase sensitive lock-in (SR844) technique, is used to create local voltage contrast of the images presented below.

\par In the regime of weak linear perturbation ($\Phi_{rf}\ll\Phi_0$; $\delta T\ll T_c$), and excluding nonlocal and nonequilibrium effects (see Ref.\cite{Gross1994RPP} and references therein), the resonant spectrum of the LSM PR$(\Phi_{app},f)$ can be modeled\cite{Zhuravel2006LTP,Zhuravel2006APL,Zhuravel2007IEEE,Zhuravel2018PRB} through additive contributions from two different origins. The first contribution is due to the shift of the resonant frequency profile $f_0(T,\Phi_{app})$ under laser illumination. As depicted in Fig.3 of Ref.\cite{Trepanier2013PRX}, a local temperature rise in a SQUID under illumination decreases the critical current $I_c$, resulting in a shift of $f_0$ at a fixed $\Phi_{dc}$. This PR component due to the $f_0$ shift is called inductive PR,\cite{Zhuravel2006APL}
\begin{equation}
\text{PR}_X \propto P_{IN}\frac{\partial|S_{21}(\Phi_{app},f)|^2}{\partial f}\delta f_0,
\end{equation}
where $P_{IN}$ is the input rf power, and PR$_X$ can be positive or negative, depending on $\Phi_{app}$ and $f$.

\par The second origin for PR is due to the change in quality factor $Q$. As the local temperature rises under the laser illumination, the dissipation increases from the quasiparticle current flowing via the tunneling resistance $R_N$ of the JJ, decreasing the $Q$ of the $|S_{21}|$ resonance dip, resulting in resistive PR,\cite{Zhuravel2006APL}
\begin{equation}
\text{PR}_R \propto P_{IN}\frac{\partial|S_{21}(\Phi_{app},f)|^2}{\partial (1/2Q)}\delta (1/2Q),
\end{equation}
which has a uniform sign as a function of frequency or flux.\cite{Zhuravel2006APL}

\par Thus the evolution of photoresponse with frequency or dc flux helps to reveal it's origins. Note that both PR$_X$ and PR$_R$ are proportional to the local rf current squared under the laser illumination, at least in the linear response regime.\cite{Zhuravel2006LTP,Zhuravel2006APL,Zhuravel2007IEEE} (See Sec.S9 of the Supp. Mat.) As a result, the photoresponse is dramatically reduced outside the resonant spectrum of the rf-SQUID.

\begin{figure}
		\includegraphics[width=0.89\columnwidth]{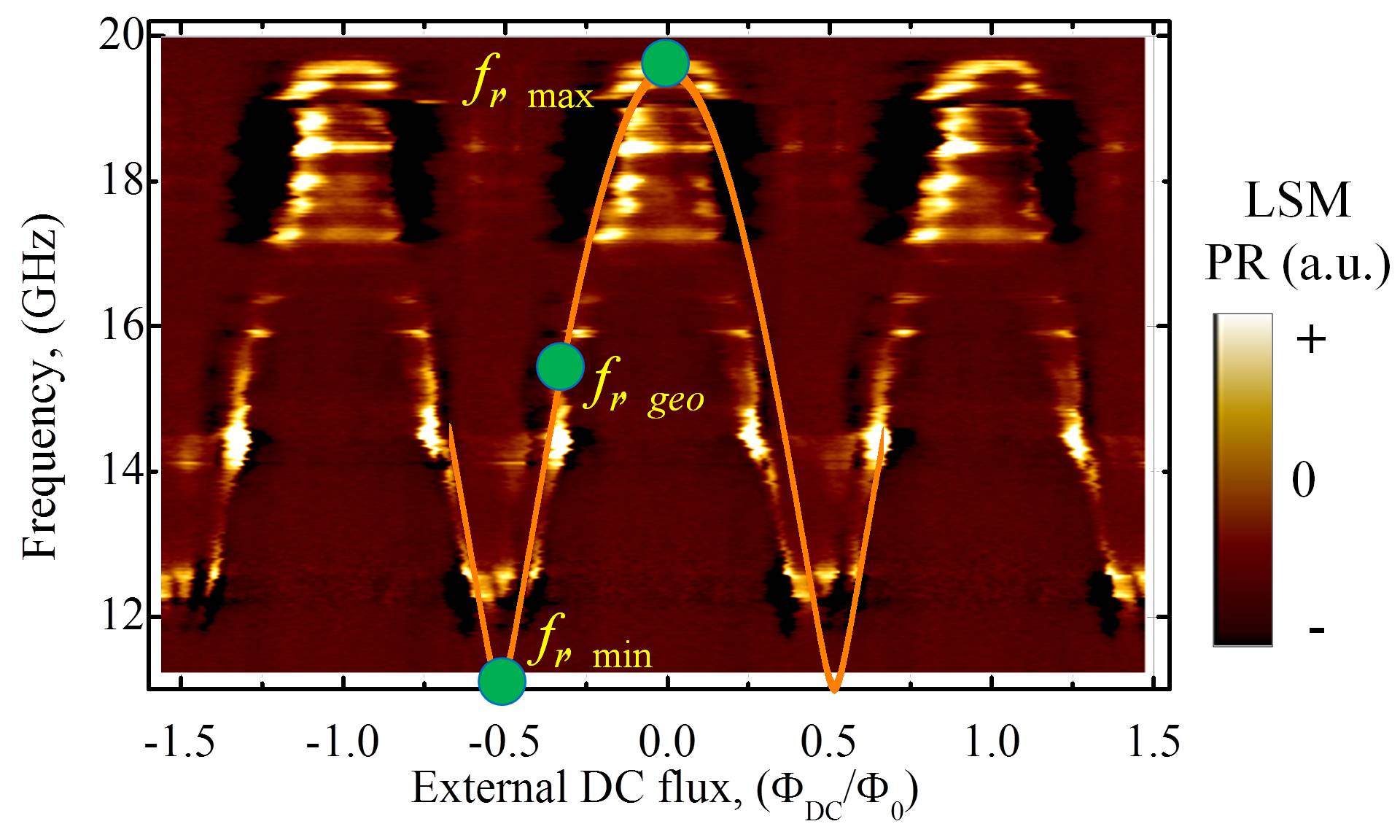}
		\caption{\label{fig:Fig2} Measured LSM photo-response of an individually probed rf-SQUID near the middle (12-th row, 14-th column) of a 27$\mathtt{\times}$27 rf-SQUID metamaterial as a function of frequency and reduced external dc flux $\Phi_{dc}/\Phi_0$ at -60 dBm rf power ($\Phi_{rf}\mathtt{\approx} 10^{-4}\Phi_{0}$) and 4.8 K.  The resonant response is outlined by the brightest areas in a false color presentation.  The orange solid line containing green circles of reference frequencies is calculated from Eq.(\ref{eq1}).\cite{Trepanier2013PRX,Trepanier2015Thesis} }
\end{figure}

\par With the LSM, we first have the opportunity to probe the \textit{global} tuning properties of the rf-SQUID metamaterial through a measurement of the \textit{local} PR from a single SQUID in the middle of the array as the globally applied $\Phi_{dc}/\Phi_0$ is varied over several periods. The resulting PR$(\Phi_{dc},f)$ from the JJ in a SQUID near the center (12-th row, 14-th column) of the array over the range of $f=10.6$ GHz (cutoff frequency of the waveguide) to 20 GHz (the maximum measurement frequency of the LSM electronics) is shown in Fig.\ref{fig:Fig2}. A clear tunability of the resonant response of the meta-atom with dc flux is evident, consistent with earlier results obtained through global transmission measurements of the entire metamaterial\cite{Trepanier2013PRX,Butz2013OptExp,Butz2013SST,Daimeng2015PRX}.  The periodic tunability of the resonant response (bright areas in Fig.\ref{fig:Fig2}) vs. magnetic flux is clearly visible with maximum value at $\Phi_{dc}= n\Phi_0$, while a minimum of $f_0$ is achieved at $\Phi_{dc}= (n+1/2)\Phi_0$, where $n$ is an integer. The strong PR features are reasonably well described by a simulated curve\cite{Trepanier2013PRX,Trepanier2015Thesis} for $f_0(T,\Phi_{app})$ from Eq.(\ref{eq1}), shown by the orange line, assuming $L_{JJ}$ varies through all values of $\cos\delta$ as a function of flux.

\par Additionally, despite observing a similar degree of tuning capability in all of the locally probed rf-SQUIDs in the array, we measured that their individual resonances are widely distributed over the full flux range. In other words, there is a significant spread of natural resonances when a small rf driving flux ($10^{-4}\Phi_0$ in this case) is applied to the rf-SQUIDs (See Supp. Mat. S5). 

\begin{figure}
		\includegraphics[width=0.89\columnwidth]{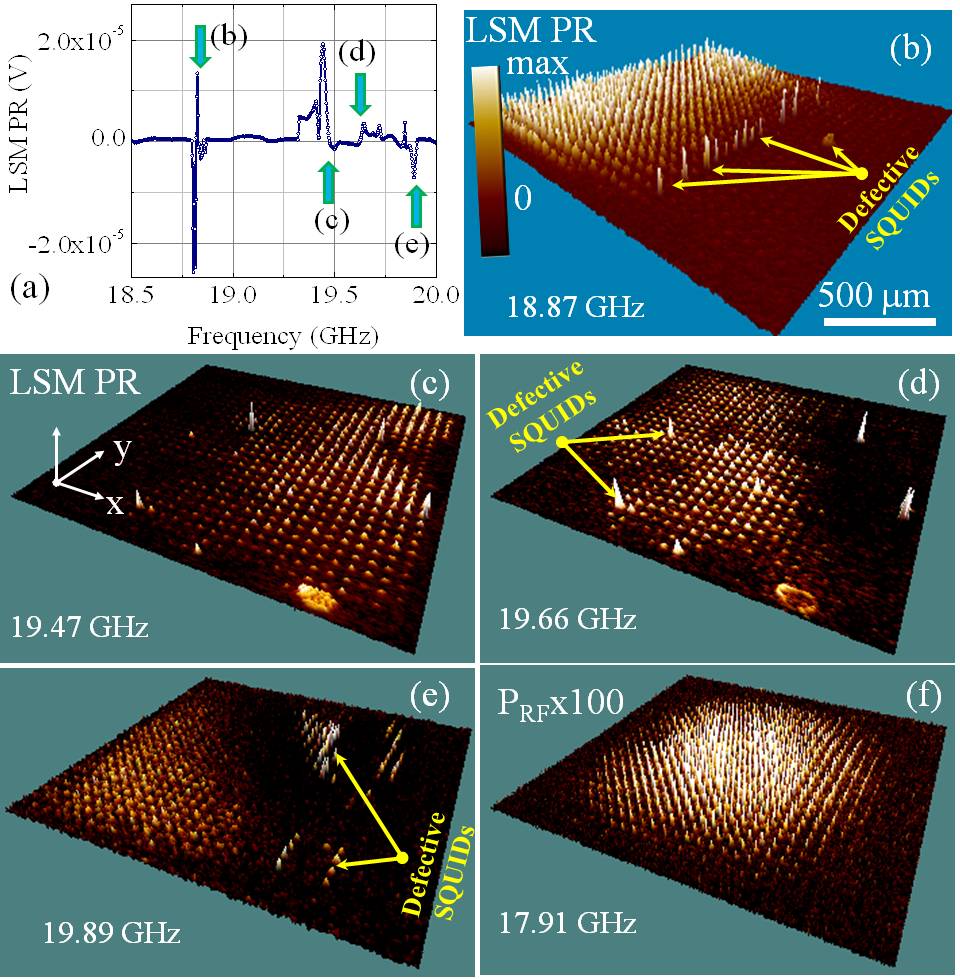}
		\caption{\label{fig:Fig3} (a) Frequency dependence of the LSM PR of an individually probed rf-SQUID located near the center of the array, at zero dc flux, -60 dBm rf power ($\Phi_{rf}\mathtt{\approx} 10^{-4}\Phi_0$), and 4.8 K.  (b)-(f) 3D LSM PR maps showing shape and position of resonant clusters at different frequencies of (b) 18.87 GHz, (c) 19.47 GHz, (d) 19.66 GHz, and (e) 19.89 GHz. (f) Pattern of PR at 17.91 GHz produced by 100 times increased rf power, $P_{in}=-40$ dBm ($\Phi_{rf}\mathtt{\approx} 10^{-3}\Phi_0$).}
\end{figure}

\par We now address the smeared out LSM PR of a single rf-SQUID in the array close to zero dc flux as shown between 17 and 20 GHz in Fig.\ref{fig:Fig2}. This PR is distinctly different from that of a single isolated SQUID (Fig.S5 of Supp. Mat.). Fig.\ref{fig:Fig3}(a) presents the experimental $\Phi_{dc} = 0$ cut of PR from Fig.\ref{fig:Fig2}. On the basis of global transmission measurement,\cite{Daimeng2015PRX,Butz2013OptExp,Trepanier2013PRX,Butz2013SST} one would naively expect PR$(\Phi_{dc}=0,f)$ to exist only at one frequency, namely $f=f_0(T,\Phi_{dc}=0)$. However, this profile of LSM PR($\Phi_{dc} = 0,f$) displays a spectrum of resonances in the wide range of 18.5 to 20 GHz, which is the range of expected magneto-inductive modes of the metamaterial\cite{Lazarides2013SST,Trepanier2015Thesis} We find that the shape of this profile is scarcely modified with variation of dc flux up to 0.2$\Phi_0$, thus we limit the spatially-resolved LSM PR investigation only to the frequency domain. Two-dimensional LSM imaging of the resonant pattern formation was done by scanning the entire area of the array by the same laser spot that was used for the local probing. A typical scanned frame consists of 600$\mathtt{\times}$600 points of PR.  Only resonating rf-SQUIDs are visible, with the PR being dominated by bright spots at the locations of the associated JJs.

\par In Fig.\ref{fig:Fig3}(b) we present LSM photo-response observed at the isolated PR resonance near 18.9 GHz (see Fig.\ref{fig:Fig3}(a)). At an input rf power of $P_{IN}$ = -60 dBm ($\Phi_{rf}\mathtt{\approx} 10^{-4}\Phi_0$), roughly half of the rf-SQUIDs show large-amplitude PR spatially grouping together in the left part of the array, with virtually no PR on the right side of array. As seen in Figs.\ref{fig:Fig3}(c)-(e), starting from $f$ = 19.45 GHz and up to 19.9 GHz, the distribution of excited SQUIDs are spatially varying as a function of driving frequency. Their structure forms stationary patterns of large clusters showing response of the rf-SQUIDs having nearly degenerate resonances. Also, LSM images in Figs.\ref{fig:Fig3}(b), (d) and (e) contain dissipative response (PR$_R$) from limited numbers of defective JJs (detail in Supp. Mat. Sec.S8).  The diagonal movement of PR structure from the upper right to the lower left corner of the array with increase of rf frequency (Figs.\ref{fig:Fig3}(c)-(e)) implies a gradient in either rf-SQUID properties, or perhaps in the dc flux applied to the metamaterial\cite{Trepanier2013PRX,Trepanier2015Thesis}.  A similar experimentally observed (but not illustrated) diagonal redistribution of the LSM PR under varied global DC flux can also be noted, and such an effect is described below.

\par A pattern with strong spatial coherence involving a majority of the rf-SQUIDs is formed by applying a stronger rf driving field at $P_{IN}$ = -40 dBm ($\Phi_{rf}\mathtt{\approx} 10^{-3}\Phi_0$) as shown in Fig.\ref{fig:Fig3}(f). One can see a dome-like distribution of strongly responding rf-SQUIDs which is expected to occur in systems of coherent rf-SQUIDs in the lowest-order magneto-inductive eigenmode of the array.\cite{Lazarides2013SST,Trepanier2015Thesis,Trepanier2017PRE} Just such a transition was observed in simulations of a 21$\mathtt{\times}$21 array as the driving amplitude activated the nonlinearity.\cite{Trepanier2015Thesis}

\begin{figure}
		\includegraphics[width=0.89\columnwidth]{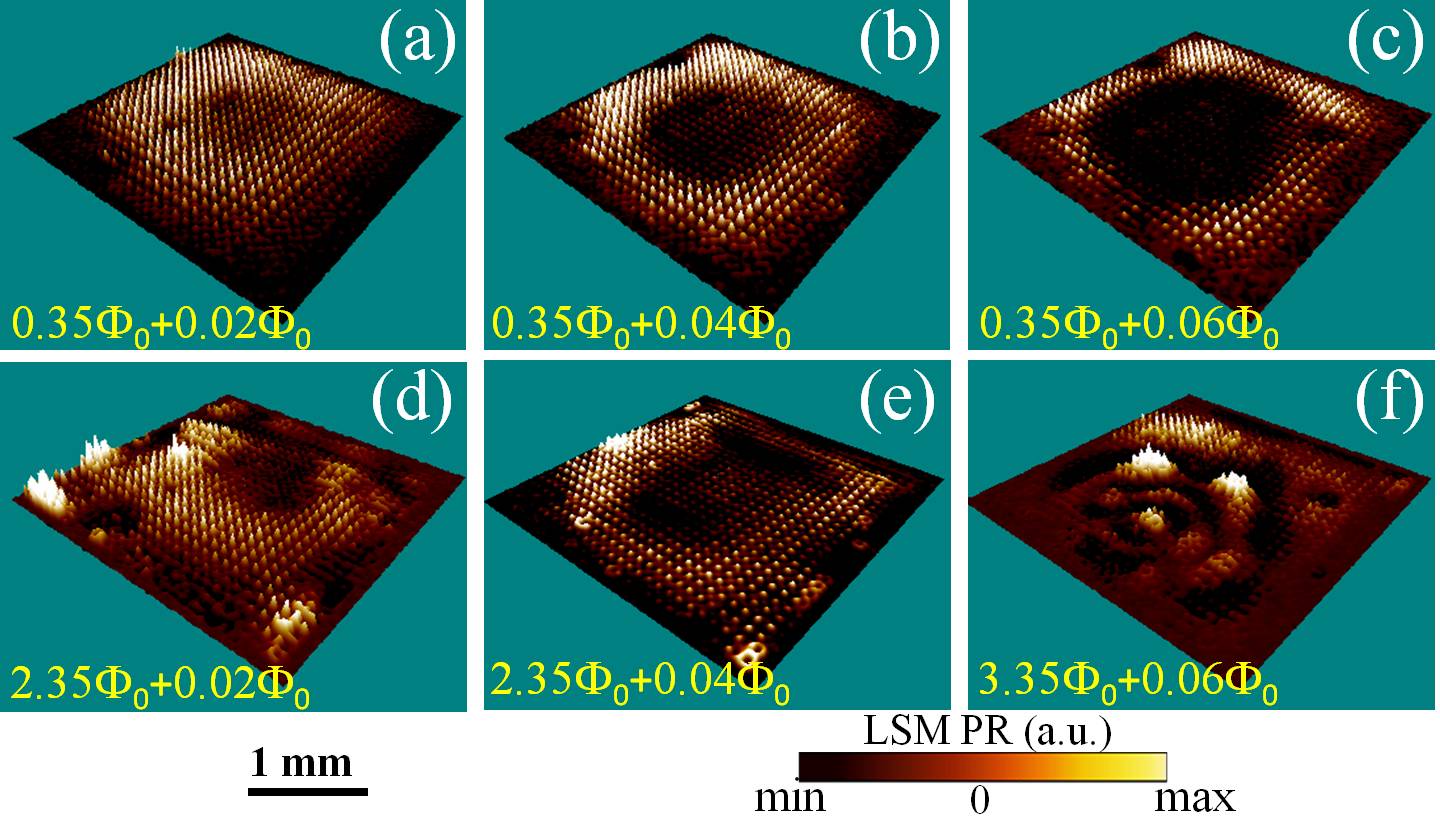}
		\caption{\label{fig:Fig4} dc magnetic flux dependent clustering of spatially coherent pattern in 27$\mathtt{\times}$27 rf-SQUIDs array that is generated by small steps of extra applied dc flux (starting from $\Phi_{dc}=0.35\Phi_{0}$) of (a) $0.02\Phi_{0}$, (b) $0.04\Phi_{0}$, and (c) $0.06\Phi_{0}$. Modification of coherent structure by integer flux quanta of (d, e) $2\Phi_{0}$ and (f) 3$\Phi_{0}$. Here, $P_{rf}=-45$ dBm ($\Phi_{rf}\mathtt{\approx} 5.6\mathtt{\times} 10^{-4}\Phi_0$) and $f=14.4$ GHz are fixed through (a)-(f).}
\end{figure}

\par The next question to examine is the influence of small variations of external dc magnetic flux on the spatial stability of the coherent state.  For this purpose we apply an external dc flux $\Phi_{dc}$=$0.35\Phi_{0}$ to obtain at an arbitrarily chosen frequency $f_0 = 14.4$ GHz $<f_{geo}$ a coherent pattern under $P_{IN}=-45$ dBm ($\Phi_{rf}\mathtt{\approx} 5.6\mathtt{\times}10^{-4}\Phi_0$) with the same distribution of LSM PR as shown in Fig.\ref{fig:Fig3}(f).  Additional DC flux increases with equal steps $\Delta\Phi_{dc}=0.02\Phi_{0}$ to explore the spatial evolution of this pattern under magnetic detuning in a flux regime showing strong tunability.  The upper row of LSM PR images in Fig.\ref{fig:Fig4} shows resonant patterns of the rf-SQUID oscillators starting from $\Phi_{dc} = 0.35 \Phi_{0} + 0.02 \Phi_{0}$ while the bottom row presents the LSM images for conjugated $\Phi_{dc}$ with an increase by 2 or 3 flux quanta.

\par The observed coherent states are clearly different from the dome-like structure in Fig.\ref{fig:Fig3}(f). The distributions in Figs.\ref{fig:Fig4}(a-c) demonstrate progressive excitation of the SQUIDs near the geometric edges of the array. The SQUIDs near the edges of the array witness a different combined dc+rf flux because they have fewer nearest neighbors than those in the center of the array.\cite{Trepanier2015Thesis,Trepanier2017PRE}  As the dc flux is tuned the edge SQUIDs come into resonance, at the expense of those in the middle of the array.

\par The na\"ive expectation is that $n\Phi_0$ flux quantization in the array will lead to repetition of the same spatial pattern at any $n$ of additional DC flux bias.  Figures \ref{fig:Fig4} (d, e) demonstrate that this is true to some extent. However, small distortions and the development of a number of additional clusters are evident compared to the structure shown in Fig.\ref{fig:Fig4}(a, b). A frequency splitting of the rf-SQUID metamaterial \textit{global} response was noted earlier at $\Phi_0$ and $2\Phi_0$.\cite{Trepanier2017PRE}  Hence we believe that these spatial features are arising from two effects of (1) increasing of DC flux gradient\cite{Trepanier2015Thesis,Trepanier2017PRE} and (2) spatial modification of resonating currents caused by defective SQUIDs. 

\par In summary, we have visualized the microscopic dark-mode states of a large nonlinear metamaterial structure using the LSM technique. The spatial variation of rf-SQUID excitation can now be elucidated while tuning driving frequency as well as dc and rf flux amplitudes. Our experiments show that the degree of coherence of the SQUID excitations is strongly enhanced for larger rf flux amplitude and is diminished by defects and inhomogeneous DC flux. 

\par See Supp. Mat. for the referenced discussion.

\par This work is supported by Volkswagen Foundation grant No. 90284, DFG grant No. US18/15, MESRF Contract No. K2-2017-081, NSF grant No. DMR-1410712, and DOE grants No. DESC0017931, DESC0018788.

\pagebreak
\pagebreak
\begin{widetext}

\newcommand{\beginsupplement}{%
		\setcounter{subsection}{0}
        \renewcommand{\thesubsection}{S\arabic{subsection}}%
        \setcounter{table}{0}
        \renewcommand{\thetable}{S\arabic{table}}%
        \setcounter{figure}{0}
        \renewcommand{\thefigure}{S\arabic{figure}}%
        \setcounter{equation}{0}
        \renewcommand{\theequation}{S\arabic{equation}}%
     }

\beginsupplement

\section*{Supplementary Material}
\subsection{Metamaterial sample under investigation}

\par In the present work, we investigate a superconducting metamaterial consisting of 27$\times$27 magnetic meta-atoms composed of rf-SQUIDs. These SQUID-based unit cells (see inset in Fig.\ref{fig:FigS1}) are equidistantly arranged in a regular square array with a lattice constant of 83 $\mu$m over a 2.5$\times$2.5 mm$^2$ area of a dielectric silicon substrate. An optical microscope image of a portion of this 27$\times$27 rf-SQUID array is shown in Fig.\ref{fig:FigS1}.

\par Nominally identical SQUIDs, each being designed as a superconducting Nb loop that is interrupted by a single Josephson tunnel junction (JJ), were manufactured using the HYPRES process utilizing 0.3 $\mu$A/$\mu$m$^2$ Nb/AlO$_x$/Nb trilayer fabrication. The individual JJ occupies a circular area with a radius $r_{JJ}$ of about 2 $\mu$m while the superconducting loop is composed of two Nb films (135 and 300 nm thick) that are galvanically connected through a Nb via. 

\par As has been discussed in Ref.\cite{Trepanier2015Thesis}, tunability is the crucial design factor of the SQUID. It allows precise adjustment of the array resonance within the measurable frequency range 10.6-20 GHz (dictated by the available waveguide) while remaining in the low noise and non-hysteretic limit ($\beta_{rf}$ is close to 0.87 for our design).\cite{Trepanier2015Thesis} The frequency can be controlled by the choice of the dimension dependent loop inductance $L_{geo}$, the overlap capacitance $C$ and the critical current $I_c$ of the JJs. In the investigated design, the area of the JJ is about 12 $\mu$m$^2$ that creates the experimentally estimated values of critical current $I_c = 2.2$ $\mu$A at temperature $T=4.2$ K corresponding to zero field Josephson inductance of about $L_{JJ} = 150$ pH. The inner and outer radii of the SQUID loops are 30 and 40 $\mu$m, respectively, establishing geometric inductance in the rf-SQUID to be $L_{geo}=120$ pH which is on the same order as $L_{JJ}$ (See Fig.\ref{fig:FigS2}). The capacitance ($C = 2.2$ pF) consists of (i) parallel plate capacitor of the overlapping area between two layers of Nb loop with SiO$_2$ dielectric and (ii) intrinsic capacitance of the Josephson junction which is almost two orders of magnitude smaller than (i). These circuit parameters control the resonant frequency of the rf-SQUID, which is tunable by dc magnetic field in the range of 10-20 GHz. More details on chip fabrication, microwave design, experimental characterization and rf simulation of almost the same array have been presented in Ref.\cite{Trepanier2015Thesis}. 

\begin{figure}
		\includegraphics[width=0.5\columnwidth]{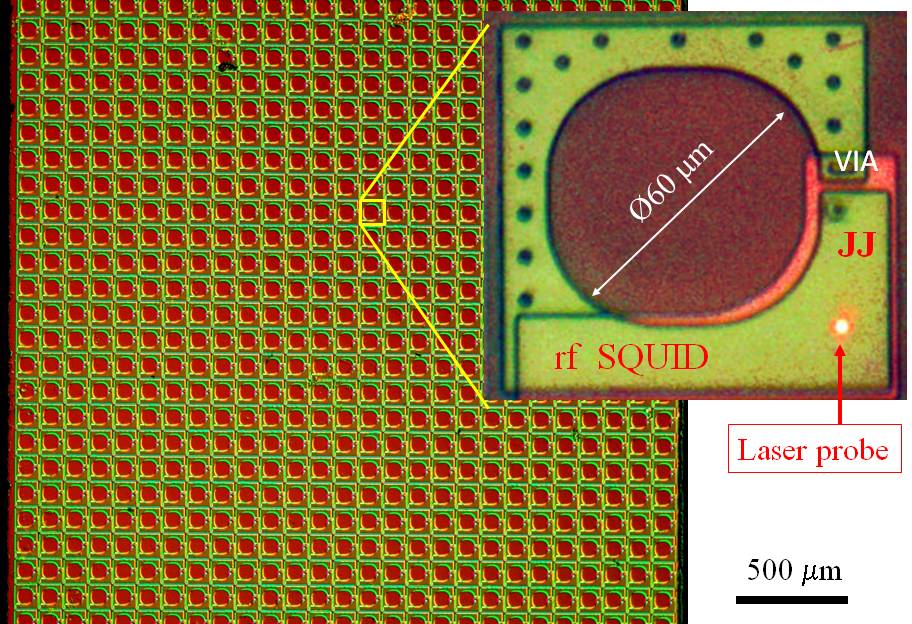}
		\caption{\label{fig:FigS1} An optical microscope image of a portion of the 27$\times$27 rf-SQUID array and a single meta-atom that composes the array. The large-scale image of the array is acquired with a polarized-light microscope while the detailed view of an individual rf-SQUID is acquired by a CCD camera incorporated in the LSM optical train. The position of the laser probe is labeled. The probe is moved to the position of the JJ while its local rf properties are probed.}
\end{figure}

\begin{figure}
		\includegraphics[width=0.5\columnwidth]{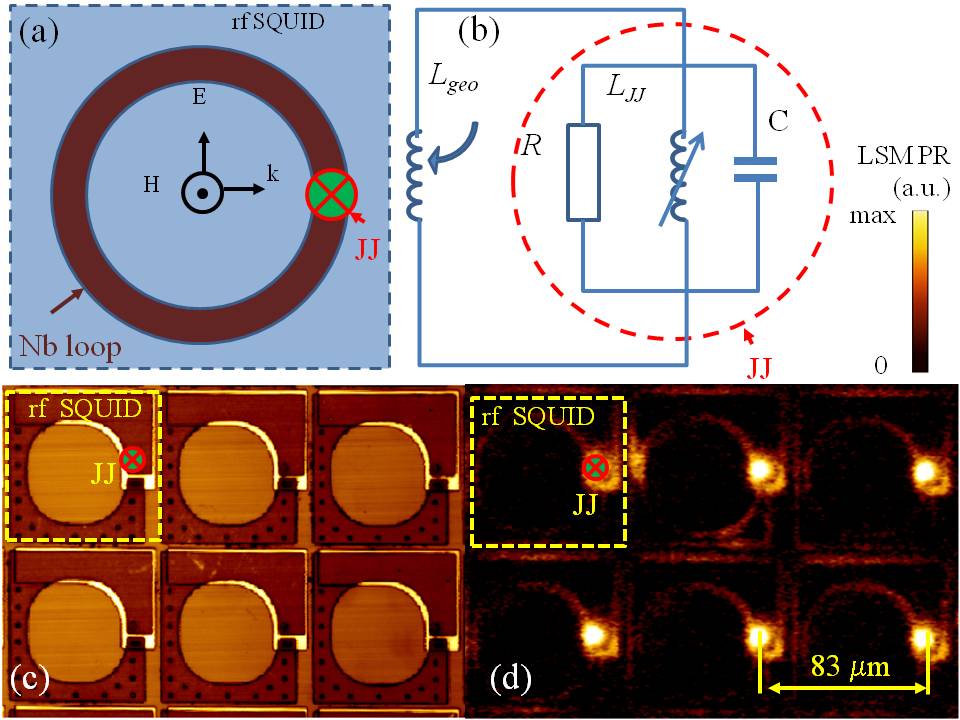}
		\caption{\label{fig:FigS2} (a) Sketch and (b) equivalent electric circuit diagram of an rf-SQUID unit cell. The red cross shows the Josephson junction (JJ). $L_{geo}$ is the geometric inductance of the Nb split ring. The red dashed circle outlines the equivalent circuit for the RCSJ model describing electrodynamics of an individual JJ in the small signal approximation. $R$ represents the equivalent resistance, $L_{JJ}$ is the Josephson inductance, and $C$ represents the capacitance between the superconducting electrodes. (c) Two-dimensional reflectivity map of 3$\times$2 unit cells acquired in the central area of a 27$\times$27 rf-SQUID array using the reflective contrast of the LSM that was obtained simultaneously with the image of the rf photoresponse. Yellow dashed squares outline an individual rf-SQUID. The red circle indicates one of six visualized JJs. Brown circumferential areas correspond to the Nb layer while black dots in the Nb layer show reflectivity from the holes serving as a flux pinning site. (d) LSM photoresponse for the six SQUIDs presented in (c).} 
\end{figure}

\par One part of the Nb loop is perforated with 3 $\mu$m diameter and 10 $\mu$m interspaced round holes to prevent the dissipative flux motion in the loop under tuned magnetic field (See Fig.\ref{fig:FigS2}). Presumably, the presence of small holes does not affect the inductance of the SQUID because the rf current densities peak on the inner edge of the loop. To make sure that this scenario is correct, we have compared two mutually complementary LSM images that were collected over the same 250$\times$170 $\mu$m$^2$ fragment of the array. As an example, Fig.\ref{fig:FigS2}(c) illustrates the LSM image of the optical reflectance $R(x,y)$ presented in brightness contrast. The detailed LSM images were acquired using an ultra-long working distance 34 mm, 20$\times$, NA = 0.42 objective lens forming a Gaussian laser probe of 1.6 $\mu$m diameter. As expected, the reflectance LSM image shows the image contrast identical to that of Fig.\ref{fig:FigS1}. The brightest areas show the geometry of surface Nb metallization (having reflectivity $R_{Nb}\sim 0.5$ at 640 nm wavelength of the scanned laser beam) while the darkest ones correspond to the uncovered fractions of Si substrate ($R_{Si}\sim 0.25$). The perforating holes are clearly distinguishable in the reflectance image. Also, the surface configuration of all of 6 spatially separated rf-SQUIDs is clearly seen inside the image. The spatially conjugated image in Fig.\ref{fig:FigS2}(d), showing the microwave contrast of photoresponse, directly verifies the expected rf current distribution of the SQUIDs. Indeed, the regions of bright circular arcs with increased rf current densities stand out at the edges of the loops while the remaining area of Nb film remains black due to almost zero response there as a result of the vanishingly small current flow. The absence of photoresponse at the edges of the perforating holes is also reliable evidence of the vortex pinning effect predicted above. What is more important, the magnitude of the response of the JJs dominates the entire response of the rf-SQUID structure.  All revealed features of spatially resolved rf response contribute to the correctness of LSM characterization of the tested sample. More detailed information on the LSM method and analyses of rf-SQUID based metamaterial is described in the next sections.

\begin{figure}
		\includegraphics[width=0.5\columnwidth]{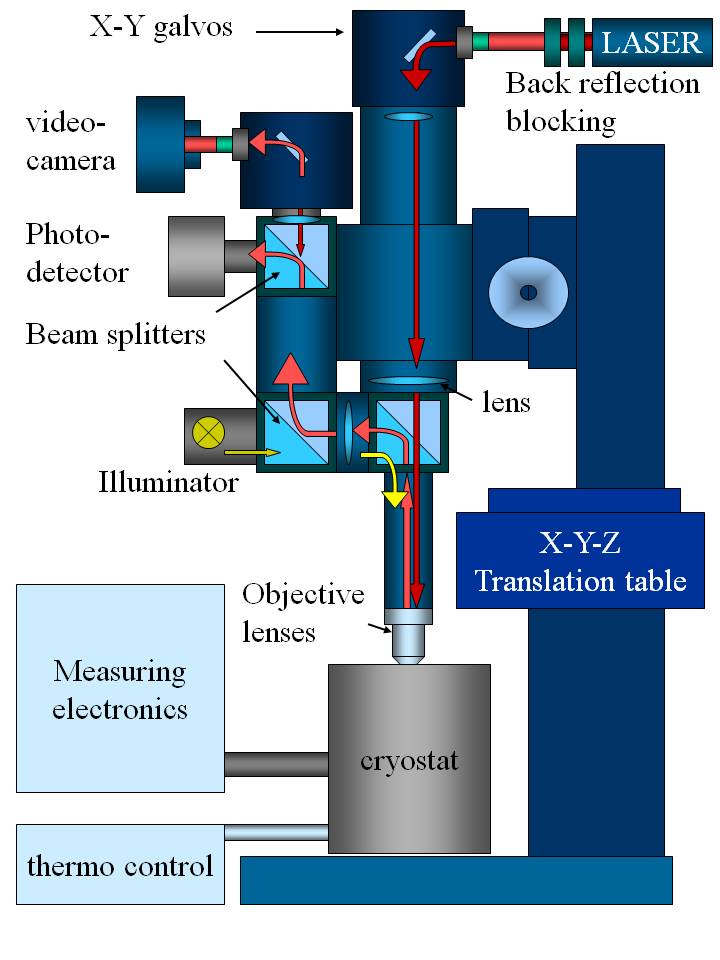}
		\caption{\label{fig:FigS3} Simplified optical train of the LSM. Tracing of probing laser beam is outlined by red while corresponding directions of reflected and direct light illumination are shown by pink and yellow arrowed lines.}
\end{figure}

\subsection{LSM technique}
\par The technique of low-temperature Laser Scanning Microscopy (LSM) has been used by us to image the in-chip position of precisely those meta-atoms that become resonating under different fixed values of dc and rf flux drive. The LSM uses the principle of point-by-point XY scanning of the superconducting planar structure by a sharply focused laser beam. Fig.\ref{fig:FigS3} shows the simplified optical train of the LSM revealing the details in Fig. 1 of the main text. 

\par Referring to Fig.\ref{fig:FigS3}, the LSM block contains the optics of a reflected light microscope equipped with a digital video-camera. The image plane of the camera is conjugated with the position of the best laser focus on the surface of the SQUID array. This gives us a handy way of adjusting the array in a proper position relative to the center point of LSM raster. While the precisely focused laser probe is scanned by XY galvanically-controlled mirrors, spatial variations of the optical properties change the intensity and polarization of the reflected beam. The reflected laser radiation is detected by a single element optical sensor as a function of probe coordinate $x$, $y$ to visualize the in-plane positions and geometry of the SQUID-based unit cells. Voltage contrast $\delta V_{opt}$ of the sensor is calibrated in units of the optical reflectance $0<R<1$ converting the $R(x,y)$ distribution into an LSM image that is almost indistinguishable from that visualized with a CCD digital camera.

\par The absorbed portion of the laser irradiation acts as a local source of heat at every point of the optical raster in the plane of the studied metamaterial. The induced perturbation is low enough not to change significantly the rf current distribution, while keeping LSM photoresponse detectable by lock-in. Note that one can ignore the nonequilibrium effects of direct Cooper pair breaking in the JJs since the long period ($\sim \mu$s) of modulation of the laser intensity compared to the short ($<$ ns) relaxation time of nonequilibrium quasi-particles.\cite{Gross1994RPP} In this case, only the heating effect of the thermally induced laser perturbation can be considered as the dominant contribution to the LSM PR. One can estimate the temperature oscillation $\delta T$ in the laser focus using the net intensity of the absorbed laser power. Taking into account a tenfold loss via LSM optics, $R_{Nb}\sim 0.5$ reflectance from Nb, and 50\% duty cycle of the amplitude modulation, the absorbed intensity is about $3.2\times 10^2$ W/m$^2$. In this case, temperature oscillations under the laser probe could be calculated as $\delta T = 2P_L/\pi a k_S=8\times 10^{-4}$ K, where $P_L$ is the incident laser heating power, $a=2l_T$ is the heat source diameter ($l_T$ is the laser modulation frequency dependent thermal healing length), and $k_S$ is the substrate material thermal conductivity: $k_S$ is of the order of 1 W/mK. A more sophisticated estimation using the linearized heat transfer coefficient between the sample and the substrate (approximately $1.1\times 10^5$ W/m$^2$K between 3 and 5 K temperature) gives the same result of $\delta T\sim 0.8$ mK. Note that the LSM photoresponse is a linear function of laser power up to $\sim 100$ $\mu$W of absorbed laser power when the critical state of the Josephson junction is achieved by overheating by laser. Temperature modulation estimates for LSM measurements have been published before in references \cite{Zhuravel2003IEEE,Zhuravel2006JSupercond,Zhuravel2010JAP,Ciovati2012RSI}

%\onecolumngrid

\begin{figure}
		\includegraphics[width=1\columnwidth]{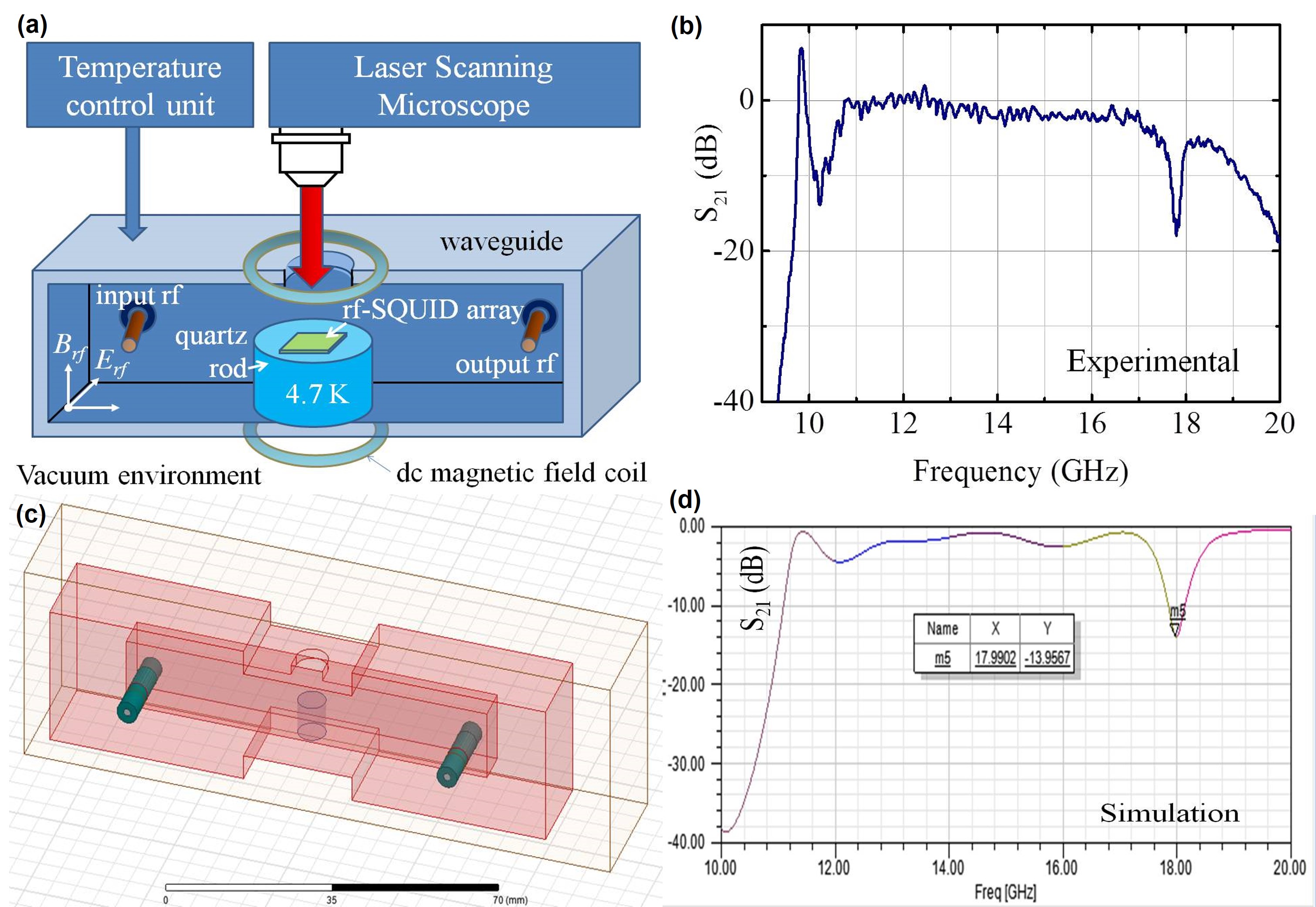}
		\caption{\label{fig:FigS4} (a) Schematic of the microwave LSM experiment. An rf-SQUID metamaterial is positioned inside a waveguide perpendicular to the rf magnetic field. The dc magnetic field is created by two superconducting coils outside the waveguide. (b) The measured frequency dependence of transmission coefficients of the waveguide cavity with a quartz rod and the SQUID sample inside at 4.8 K. The input power is $-60$ dBm ($\Phi_{rf}\approx 10^{-4}\Phi_0$) while the output rf signal is 59 dB amplified by two external amplifiers connected in series. (c) Sketch of the waveguide cavity assembly which has been used for HFSS simulation. The waveguide section is terminated with the two waveguide-to-coaxial SMA connectors directing the rf signal in and out of the waveguide. (d) HFSS simulated frequency dependence of transmission coefficients of the waveguide with a quartz rod and the SQUID sample inside. }
\end{figure}
%\twocolumngrid

\subsection{Cryogenic rf waveguide with optical access}
\par Several technical enhancements have been made to adapt the existing LSM technique to the specific conditions suitable for experiments with rf-SQUIDs. First, as shown in Fig.\ref{fig:FigS4}(a), the array of 27$\times$27 SQUIDs is placed inside a rectangular Ku band-like waveguide whose dimensions are reduced from 15.8$\times$7.9 mm$^2$ (standard WR62 waveguide) to 14$\times$7 mm$^2$. The length of the waveguide is 80 mm. The waveguide to coaxial coupler design was optimized with ANSYS HFSS electromagnetic simulation software in order to avoid standing waves at the sample location. The main requirement of such optimization was to provide a uniform rf magnetic excitation over the area of SQUID array. This was achieved by positioning the moderate size ($d << \lambda$) SQUID array in a vertical plane of symmetry of the rectangular waveguide, operating in the regime of a single propagating TE$_{10}$ wave (with a low level of standing waves). In a rectangular waveguide, the magnetic field component of the fundamental TE$_{10}$ mode is predominantly perpendicular to the vertical plane of symmetry. A similar type of waveguide was recently used for a study of the collective response of a SQUID array in Ref.\cite{Averkin2017IEEE} Note that the HFSS simulated (see Fig.\ref{fig:FigS4}(d)) and measured (see Fig.\ref{fig:FigS4}(b)) transmission through the empty waveguide demonstrate almost identical features. In particular the microwave transmission properties of the waveguide are not severely compromised by the presence of the quartz rod or hole in the wall of the waveguide.

\par Secondly, the effective cooling below $T_c$ of the SQUIDs has been achieved in the vacuum environment of the optical cryostat. It was organized by gluing the array to a quartz rod by low-temperature (Apiezon N) vacuum grease providing reliable thermal contact with the Cu waveguide. Thirdly, the substrate was exactly positioned in a homogenous rf field in such a way that the orientation of the dominant magnetic component was orthogonal to the XY plane of the SQUIDs inside the waveguide. Additionally, a set of superconducting Helmholtz coils outside the waveguide create a homogenous dc magnetic flux through the array. Finally, the entire structure was surrounded by mu-metal magnetic shield and the temperature was stabilized in the range between 4.5 and 10 K with an accuracy of 1 mK. Both the mu-metal and Cu waveguide have coaxial through holes to allow entry of the laser beam into the waveguide (Fig.\ref{fig:FigS4}(a),(c)). 

\subsection{LSM response of individual rf-SQUIDs: precise adjustment of the resonance to the desired frequency}
\begin{figure}
		\includegraphics[width=0.5\columnwidth]{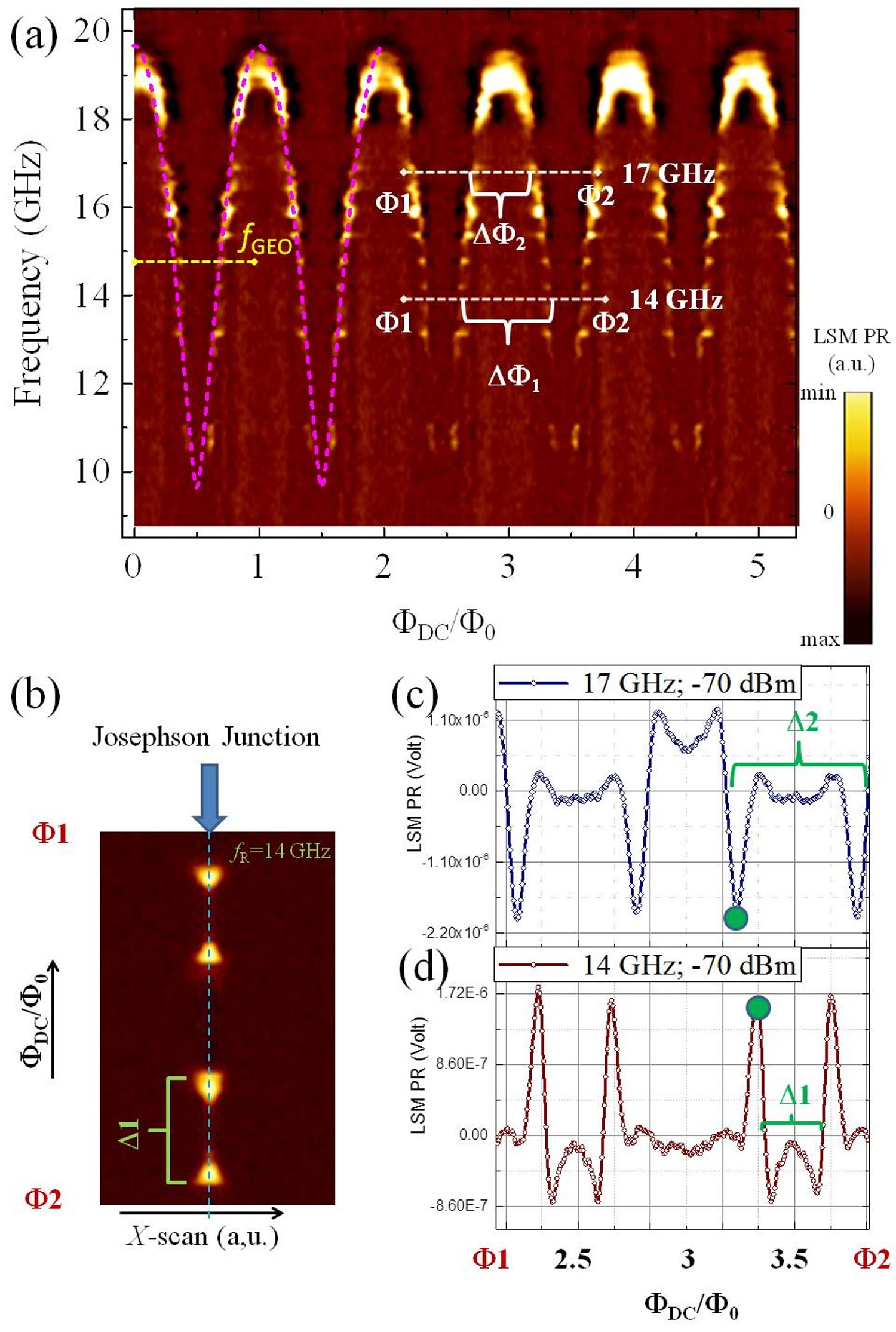}
		\caption{\label{fig:FigS5} (a) Measured LSM photoresponse of an individual isolated rf-SQUID (that has been made with the same process which was used to manufacture the 27$\times$27 rf-SQUID array) as a function of frequency and reduced external dc flux $\Phi_{dc}/\Phi_0$ at $-70$ dBm rf power ($\Phi_{rf}\approx 3.2\times 10^{-5}\Phi_0$). The resonant response is outlined by the brightest areas in a false color presentation. Two dashed lines along the $\Phi_{dc}/\Phi_0$ coordinate at 14 GHz and 17 GHz shows line cuts for corresponding (d) and (c) profiles. Note that inverted LSM contrast is used relative to the one used in Fig.2 of the main text. (b) Combined spatial and flux line scan through the JJ of the single SQUID.}
\end{figure}

\par Fig.\ref{fig:FigS5}(a) shows LSM photoresponse obtained by dwelling the laser beam on a Josephson junction which belongs to a single isolated rf-SQUID being designed as a test structure. It is manufactured using the same HYPRES process used to make the 27$\times$27 array. The periodic tunability of the resonant response (bright areas in Fig.\ref{fig:FigS5}(a)) vs. applied perpendicular dc magnetic flux is evident, making a complete analogy with the results of Fig. 2(a) in the main text. Moreover, the positions of maxima in PR are also described well by Eq.(1), as shown by the dashed magenta line. As evident from Fig.\ref{fig:FigS5}(a)), this resonant portrait of the LSM PR of the rf-SQUID of the fundamental ($f_0$) mode is uniquely defined by local peak values of LSM PR$(x_0, y_0; \Phi_{dc}, \Phi_{rf})$, where the static localized laser probe $(x_0, y_0)$ is positioned on the Josephson junction in the examined rf-SQUID. It is important to keep in mind that this photoresponse is strongly reduced outside the resonant spectrum of the rf-SQUID. This feature can be applied to find the proper magnitude of the external dc flux that tunes the resonant response of the spatially selected rf-SQUID exactly to the frequency of required resonance. This was done by using fixed-frequency LSM probing over a range corresponding to $\Phi_{ext}$ between $\Phi_{1}$ and $\Phi_{2}$ that are centered around $3\Phi_0$ as shown by the white dashed lines in Fig.\ref{fig:FigS5}(a). The experimental values of $f_0$ at 14 (and 17) GHz were fixed arbitrarily to be a little below (and above) the geometric resonance $f_{geo} = 14.8$ GHz (see corresponding positions in Fig.\ref{fig:FigS5}(a)). As evident from Fig.\ref{fig:FigS5}(c),(d), the resonances of the rf-SQUID appear as well separated peaks of the LSM PR$(x_0, y_0; \Phi_{ext})$ along the $\Phi_{dc}/\Phi_0$ coordinate. Almost all $\Phi_{dc} = n\Phi_0$ peaks are equidistantly ($\pm 0.5 \Delta\Phi_{1,2}$) positioned relative to integer values of field flux coordinates scaled in dimensionless units of $n=\Phi_{dc}/\Phi_0$. As expected, further increasing of dc magnetic field multiplies the double-peak structure of LSM PR($\Phi_{dc}/\Phi_0$) periodically with $n$ and a periodicity of $\Phi_0$. The effect is universal relative to $\Phi_{dc}$ dependent frequency tuning and can be used for precise adjustment of the rf-SQUID response to any desired frequency. Note that distance $\Delta\Phi$ (i.e. $\Delta\Phi_1$ and $\Delta\Phi_2$ in shown case) between peaks of LSM PR($\Phi_{dc}/\Phi_0$) is correlated exactly with the position of rf-SQUID resonances in the frequency domain. Therefore, one can expect that if no variations in LSM PR($\Phi_{dc}/\Phi_0,f$) portrait are appearing while refocusing the laser probe on other SQUIDs, this would demonstrate their identical natural resonances in the 2D array. On the other hand, variability of LSM PR($\Phi_{dc}/\Phi_0,f,x_0,y_0$) portraits in different $x_0$, $y_0$ positions of the laser beam gives direct evidence of spreading of the natural resonances of individual rf-SQUIDs over the 2D array. The effect can be utilized for microwave characterization of an rf-SQUID array as demonstrated in the next section.

\par It is clear that analyzing the entire set of LSM PR($\Phi_{dc}/\Phi_0,f,x_0,y_0$) data requires a large investment of time to collect detailed information from all of 27$\times$27 = 729 rf-SQUIDs. The problem is solved much faster if the dc flux line cuts of Fig.\ref{fig:FigS5}(c),(d) are concatenated spatially while scanning through the JJ in a SQUID. Fig.\ref{fig:FigS5}(b) shows a simple example of a spatial scan like Fig.\ref{fig:FigS5}(d) that details the spatial modification of this profile across and close to a single JJ. The image was obtained by repeating 1D scans along the $x$ axis that contains a JJ while dc flux was incrementally changed by $10^{-3}\Phi_0$ steps between $\Phi_1$ and $\Phi_2$ at $f=14$ GHz. The brightest areas Fig.\ref{fig:FigS5}(b) again show the sharp features at $\Delta\Phi_1$ separated resonances, but also exhibits fine structure along the spatial scan direction. Note that almost the same procedure can be used for spatially resolved LSM imaging of the distributed SQUID resonances in the frequency domain by repeating $x$-scans at fixed magnetic field as the rf frequency changes. Both procedures are complementary and can be used to investigate spatial spreading of microwave resonances in the planar metamaterial. 

\subsection{Dispersion of natural resonances in the rf-SQUID metamaterial}
\par Fig.\ref{fig:FigS6}(a) demonstrates an example of the inhomogeneous dc flux response of the natural resonances that are observed in the central area of the 27$\times$27 rf-SQUID array under small rf drive at $P_{rf}=-70$ dBm ($\Phi_{rf}\approx 3.2\times 10^{-5}\Phi_0$). This dc flux line scan through multiple SQUIDs is similar to that in Fig.\ref{fig:FigS5}(b). In Fig.\ref{fig:FigS6}(a), one can see that even neighboring SQUIDs are effectively decoupled and their resonant frequencies (at a given value of $\Phi_{dc}/\Phi_0$) vary significantly within the array. However, as shown in Fig.\ref{fig:FigS6}(b), the spread of resonances is greatly reduced under strong rf driving amplitude at $P_{rf}=-40$ dBm ($\Phi_{rf}\approx 10^{-3}\Phi_0$). One can see that different SQUIDs resonate at almost the same values of the threaded dc flux.

\begin{figure}
		\includegraphics[width=0.6\columnwidth]{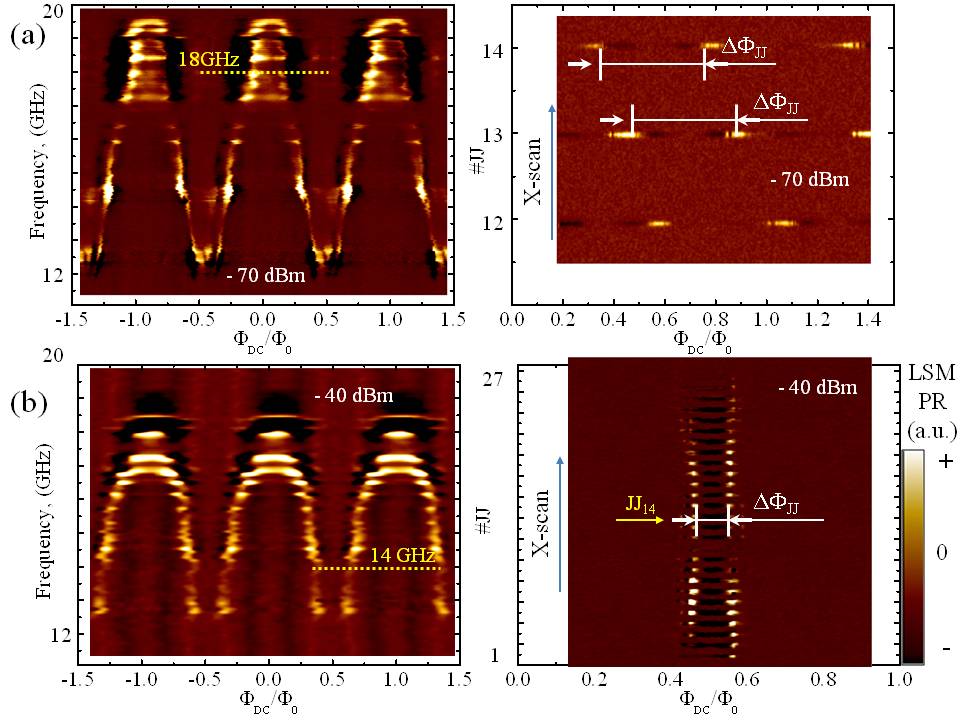}
		\caption{\label{fig:FigS6} Spatial 1D distribution of the resonances (bright areas) vs. applied DC flux at $f=17.9$ GHz for (a) three SQUIDs at $P_{rf}=-70$ dBm ($\Phi_{rf}\approx 3.2\times 10^{-5}\Phi_0$) and (b) 27 SQUIDs at $P_{rf}=-40$ dBm ($\Phi_{rf}\approx 10^{-3}\Phi_0$) measured by LSM along the 14-th row of the 27$\times$27 rf-SQUID array that is described in the main text. }
\end{figure}

\par A similar distribution of spatially varying resonances is illustrated in Fig.\ref{fig:FigS7}(b) that was LSM acquired through the same $x$-scan across the array in a frequency domain imaging mode. Again, variations of resonant frequencies are clearly visible and plotted as corresponding profiles of LSM PR($f, x_0, y_0$) at fixed $\Phi_{dc}/\Phi_0\mathtt{\sim}0.45$  and $P_{rf}= -60$ dBm ($\Phi_{rf}\approx 10^{-4}\Phi_0$) in Fig.\ref{fig:FigS7}(a).

\begin{figure}
		\includegraphics[width=0.65\columnwidth]{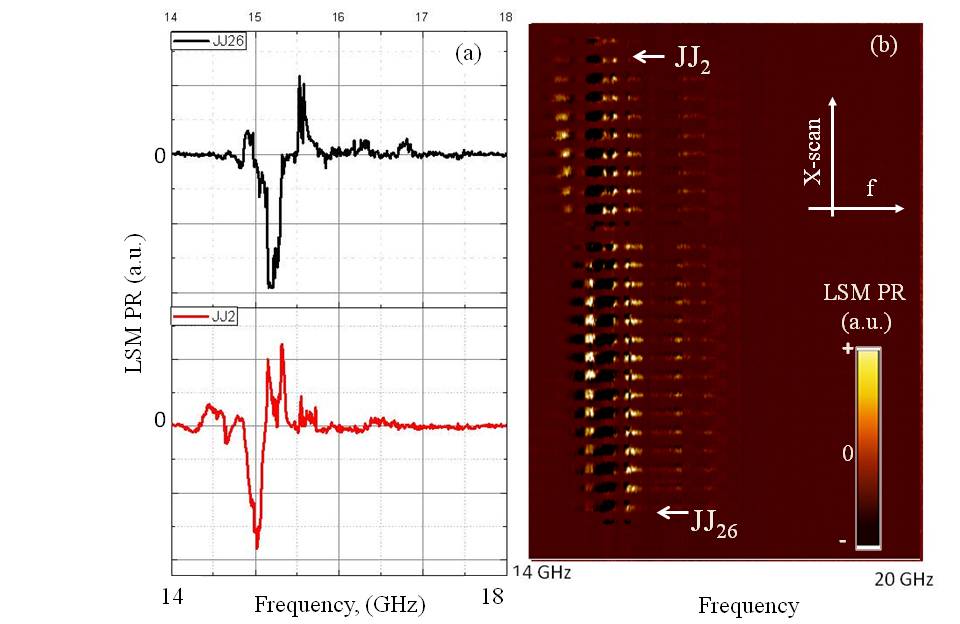}
		\caption{\label{fig:FigS7} (a) Profiles of the LSM PR vs. frequency at fixed $\Phi_{dc}/\Phi_0\mathtt{\sim}0.45$  and $P_{rf}= -60$ dBm ($\Phi_{rf}\approx 10^{-4}\Phi_0$) for two SQUIDs containing Josephson junctions JJ2 and JJ26 according to the numeration that is indicated in (b). (b) shows a line scan through the 14-th row of the 27$\times$27 rf-SQUID array, as a function of frequency between 14 and 20 GHz. }
\end{figure}

\subsection{DISPERSION OF NATURAL RESONANCES IN THE RF-SQUID METAMATERIAL: 2D verification}
\par Fig.\ref{fig:FigS13} shows frequency dependent modification of the resonance structure in whole area of 27$\times$27 rf-SQUIDs array at $P_{rf}=-70$ dBm ($\Phi_{rf}\approx 3.2\times 10^{-5}\Phi_0$). The frequency interval of rf excitation is chosen so that almost all SQUIDs in the array become resonating inside this interval. One can see that the formation of large-scale coherent clusters has been in progress between 19.4 GHz and 19.9 GHz showing a lower limit for dispersion of natural resonances of about 0.5 GHz. The isolated resonances of individual SQUIDs, are visible in the LSM images below 19.4 GHz and above 19.9 GHz, and result from the defective JJs increasing dispersion of natural resonances up to a higher limit of about 1.2 GHz. However, the population of the “defected” SQUIDs does not exceed 3\% of the total number of 27$\times$27 rf-SQUIDs in the array.

\begin{figure}
		\includegraphics[width=0.6\columnwidth]{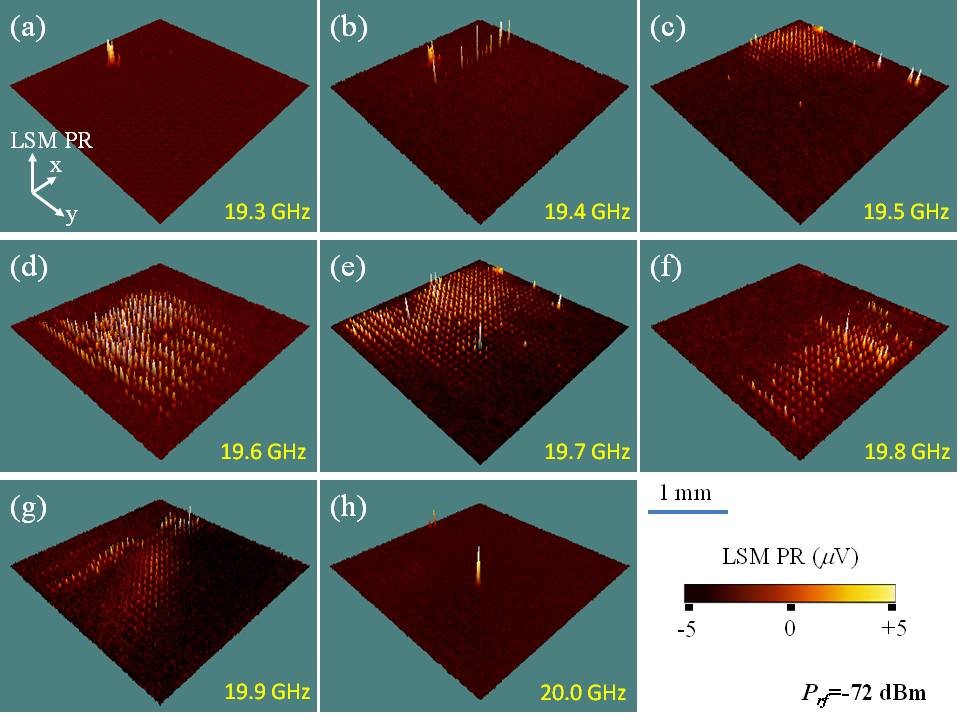}
		\caption{\label{fig:FigS13} Frequency dependence of the LSM photoresponse of resonating meta-atoms in area of 27$\times$27 rf-SQUID array at $T=4.8$ K and $P_{rf}=-70$ dBm ($\Phi_{rf}\approx 3.2\times 10^{-5}\Phi_0$). The (a-h) series is acquired with 0.1 GHz frequency steps in the vicinity of fundamental resonance of about $f_0=19.6$ GHz at zero DC flux.   }
\end{figure}

\begin{figure}
		\includegraphics[width=0.5\columnwidth]{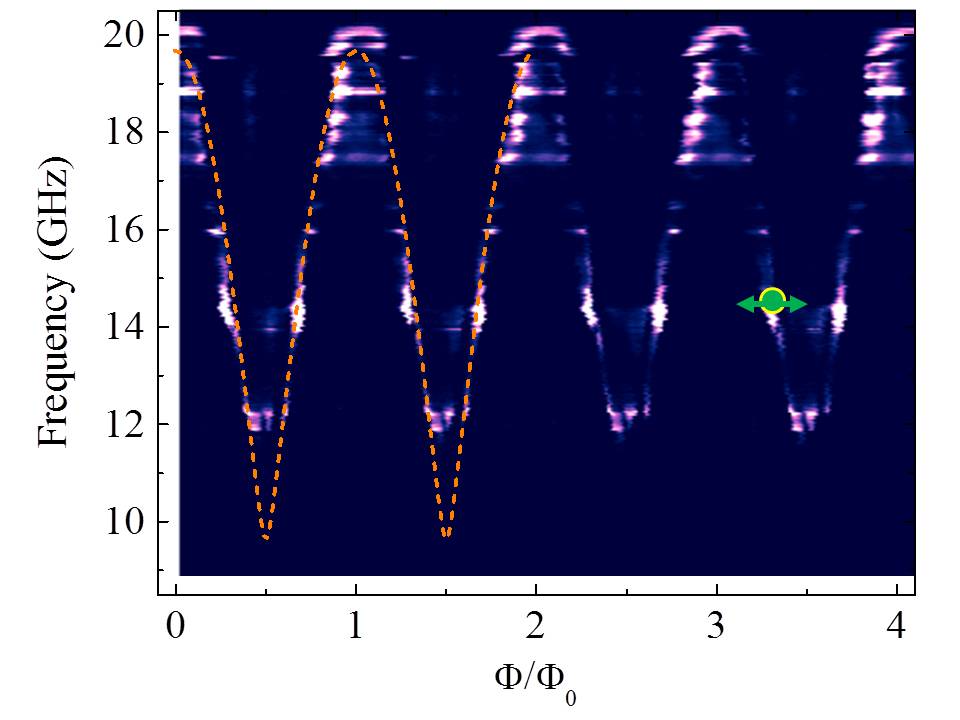}
		\caption{\label{fig:FigS8} Measured LSM photoresponse of an individually probed rf-SQUID near the middle (12-th row, 14-th column) of a 27$\times$27 rf-SQUID metamaterial as a function of frequency and reduced external dc flux $\Phi_{dc}/\Phi_0$ at -60 dBm rf power ($\Phi_{rf} \approx 10^{-4}\Phi_0$) and 4.8 K. The resonant response is outlined by the brightest areas in a false color presentation. The arrowed green line shows the direction of $\Phi_{dc}$ tuning close to $\Phi_{dc}=3.35\Phi_0$ that was used to obtain the LSM images of Fig.\ref{fig:FigS9}. }
\end{figure}

\begin{figure}
		\includegraphics[width=0.5\columnwidth]{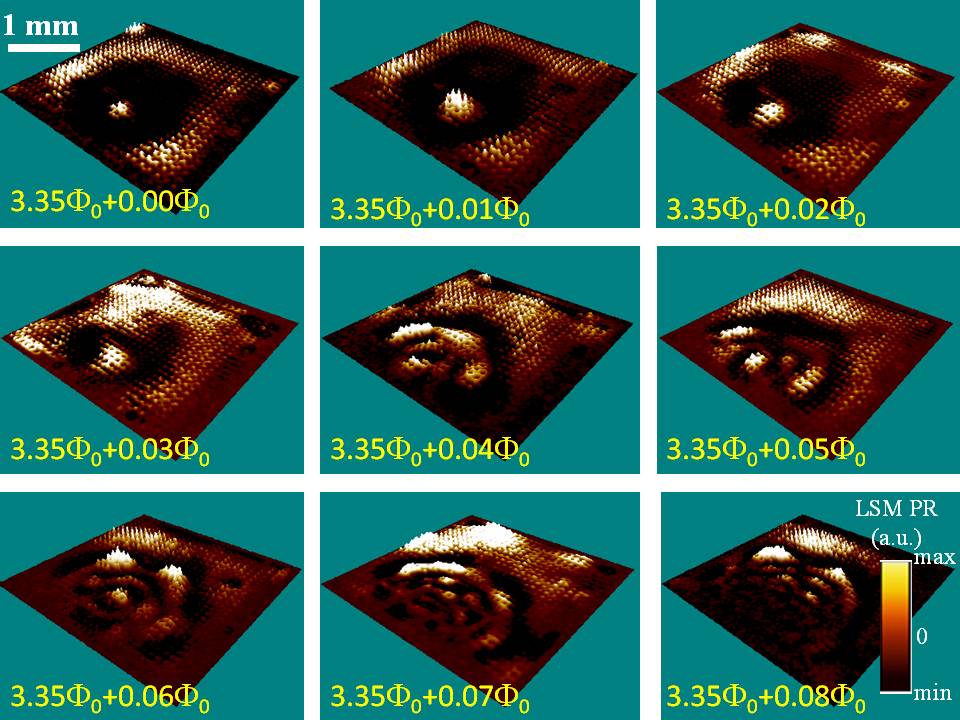}
		\caption{\label{fig:FigS9} dc magnetic flux dependent modification of resonant pattern in 27$\times$27 rf-SQUIDs array that is generated by $0.01\Phi_0$ steps of extra applied dc flux (starting from $\Phi_{dc}=3.35\Phi_0$. Here, $P_{rf} = -45$ dBm ($\Phi_{rf}\approx 5.6\times 10^{-4}\Phi_0$), $T=4.8$ K and $f = 14.4$ GHz are fixed through all images. }
\end{figure}

\subsection{High dc field}
\par Fig.\ref{fig:FigS8} repeats the data presented in Fig. 2 of the main text. However, the range of $\Phi_{dc}/\Phi_0$ is increased up to 4 to show the invariance of the measured LSM photoresponse of individually probed rf-SQUIDs in a wide range of external dc flux tuning. Additionally, the evident locality of the LSM response is demonstrated when all features are similar to the response of a single SQUID (see Fig.\ref{fig:FigS5} for comparison). I.e. no phantom-like images are generated in Fig.\ref{fig:FigS8} despite the inhomogeneous surrounding population of SQUIDs. However, such nearly ideal behavior reflects (most likely) only the trend but not the real collective response of the individually probed SQUID. This collective behavior looks more complicated and undergoes a gradual change in spatial structure of microwave response with dc flux. Fig.\ref{fig:FigS9} shows the spatial rearrangement of the resonant pattern in the 27$\times$27 rf-SQUIDs array that is generated by $0.01\Phi_0$ steps of extra applied dc flux (starting from $\Phi_{dc}=3.35\Phi_0$. Here, $P_{rf} = -45$ dBm ($\Phi_{rf}\approx 5.6\times 10^{-4}\Phi_0$), $T=4.8$ K and $f = 14.4$ GHz remain fixed through all images, as in Fig.4(a)-(c).

\subsection{Healing of defects}
\par Fig.\ref{fig:FigS10} shows that the influence of a defective SQUID has a very limited range due to the effective healing of the synchronous response outside an area of only a few surrounding SQUIDs. Fig.\ref{fig:FigS10}(b) shows an almost homogenous distribution of synchronously responsive SQUIDs in area 1 (see Fig.\ref{fig:FigS10}(a)) that generates large LSM PR of the corresponding JJs (brightest spots of equal amplitude). The position of a defective SQUID is outlined by area 2 in Fig.\ref{fig:FigS10}(a) and its detailed view is imaged as a black circular pattern of inverted PR contrast in Fig.\ref{fig:FigS10}(c). One can see that the virgin state of synchronously responsive SQUIDs is restored in the region beyond one to two unit cells.

\begin{figure}
		\includegraphics[width=0.5\columnwidth]{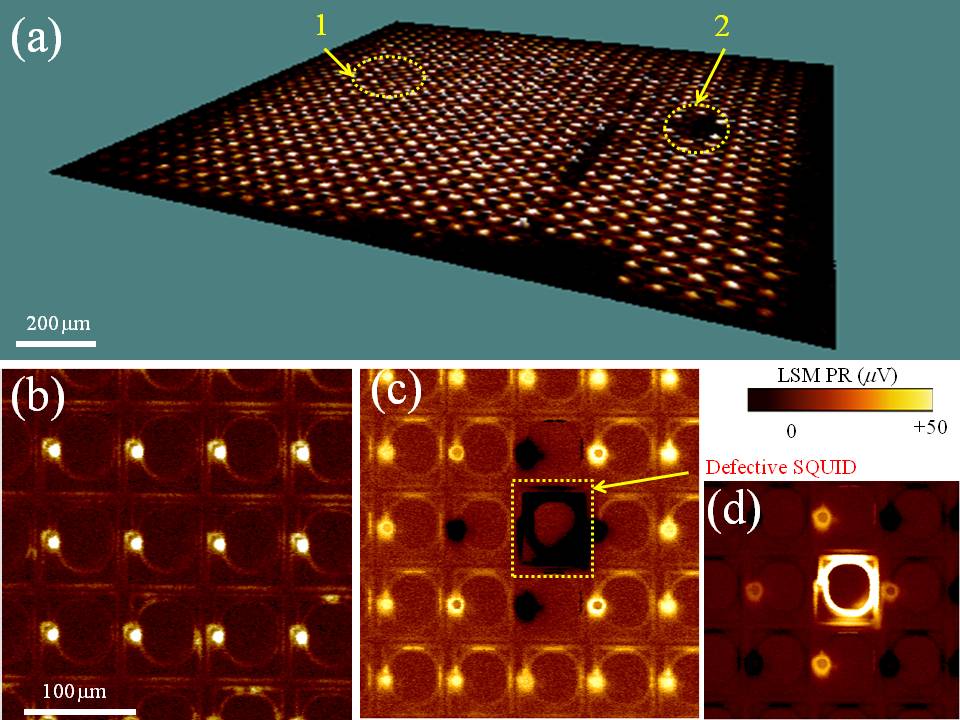}
		\caption{\label{fig:FigS10} (a) Whole-area frequency synchronized response of meta-atoms in a 27$\times$27 rf-SQUID metamaterial at 19.84 GHz together with detailed views (b) defect free area and (c) area with a single defective SQUID at $-30$ dBm rf power ($\Phi_{rf}\approx 3.2\times 10^{-3}\Phi_0$) and 4.8 K. (d) shows part of (c) with inverted contrast. }
\end{figure}

\subsection{Microwave power dependent LSM PR of a single rf SQUID: Examination of the extent of the linear response approximation }
\par In many aspects, the LSM PR of a single rf SQUID-based oscillator can be approximated in the frame of a phenomenological linearized model (see Equations 2 and 3 in the main text) which was used to describe rf response of the classical superconducting split-ring resonator. The difference in modeling PR from a rf-SQUID structure compared to that of a split-ring resonator is that the lumped capacitance of the split-ring resonator is replaced by a single Josephson junction of the SQUID structure. In the latter case the limits of linear response are not well defined due to permanent tuning of SQUID resonance under modification of Josephson inductance by a variable rf drive. In the case of high-quality JJs with a sufficiency homogenous tunneling conductivity, $\sigma_n(x,y) = 1/R$, the bolometric modulation of Josephson inductance underneath the laser probe
\begin{equation}
\delta L_{JJ}(x_0,y_0,\delta T) = \left[ d\left( \frac{\Phi_0}{2\pi I_c \cos \phi_0}\right)/dT \right]\delta T
\end{equation}
can be associated with local variations of the pair tunneling current density\cite{Gross1994RPP}
\begin{equation}
\delta I_s(x,y) = \delta I^{jj}_s(x,y) + \delta I^\phi_s(x,y),
\end{equation} 
where $(x,y)$ is the location of the laser beam focus on the sample surface; $\delta I^{jj}_s$ and $\delta I^\phi_s$ are, respectively, the change of the pair tunneling current due to the local change of the maximum pair tunneling current density and the local phase difference. At small perturbation ($\delta J_S/J_S \ll 1$) to the JJ from the laser beam, the local heating of laser probe onto the JJ varies the pair tunneling current and produces the LSM PR$_{JJ}\propto (\partial J_c / \partial T)\sin\phi$ as a result of localized reduction of $J_c$ underneath the thermal spot that oscillates in temperature synchronously with the laser modulation.

\par Fig.\ref{fig:FigS11} demonstrates advantages of LSM procedure that allows verification of linearity of LSM PR in the function of tunneling JJ current excluding the effect of varied $\phi(P_{rf})$. There in Fig.\ref{fig:FigS11}(a) and (b), the LSM PR($\phi,P_{rf}$) is probed by a stationary laser beam that is focused directly on a JJ. Note that we exclude the effect of inductive coupling from consideration by using only a single isolated SQUID structure on the substrate with the same dimensions and properties as those in the 27$\times$27 SQUIDs array. The in-plane size of the probed SQUID was also three times multiplied increasing its filling factor inside the waveguide while keeping the same HYPRES process of 0.3 $\mu$A/$\mu$m$^2$ Nb/AlO$_x$/Nb trilayer fabrication. For this SQUID, $\Phi_{rf}=0.006 \Phi_0$ at -35 dBm in units of rf power as mentioned in Fig.\ref{fig:FigS11}.

\begin{figure}
		\includegraphics[width=0.6\columnwidth]{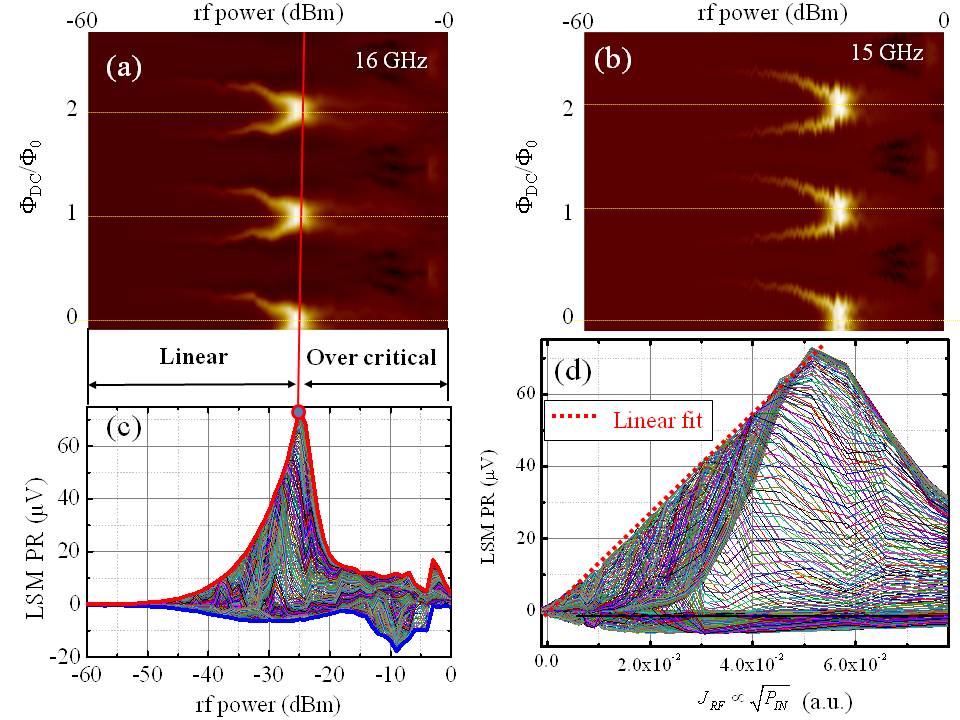}
		\caption{\label{fig:FigS11} Measured LSM photoresponse of a single isolated rf-SQUID as a function of external rf power ($P_{in}$) and reduced dc fluxes ($\Phi_{DC}/\Phi_0$) at different fixed frequencies of (a) 16 GHz and (b) 15 GHz. The resonant response is outlined by the brightest areas in a false color presentation. The plots of LSM PR in terms of (c) logarithmically scaled rf power and (d) linearly scaled rf flux ($\Phi_{rf}\sim \sqrt{P_{in}}$). The red dashed line in (d) is a linear fit to the data representing positive peaks of the LSM PR. }
\end{figure}

\par The main idea of the proposed experiment is to combine the usual recording of rf power dependence with continuous magnetic sweeping of the Josephson phase difference, allowing us to explore any desired resonant frequency of the SQUID. This universality is indicated in Fig.\ref{fig:FigS11}(a),(b) where the rf power dependence of the LSM PR is presented for two different resonances at 15 and 16 GHz. The brightness codes of the LSM PR of Fig.\ref{fig:FigS11}(a) was converted in Fig.\ref{fig:FigS11}(c) into a set of superimposed LSM PR($P_{rf}$) profiles, each of which has been measured at fixed $\Phi_{dc}$ with $\Phi_0/300$ separation. A monotonic rise of positive maximum of the LSM PR with increasing rf power up to $-25$ dBm is clearly evident. This originates from enhancement of quasi-particle tunneling which gives the most dissipative contribution to LSM PR. The subsequent degradation of the photoresponse beyond $-25$ dBm arises from an overcritical rf drive through the JJ. Another feature to note is a small inductive contribution of Cooper pairs tunneling that is visible in Fig.\ref{fig:FigS11}(c) as a change of negative minimum of the LSM PR below $-20$ dBm. We did not use the procedure of PR component partition here to show this effect in detail.

\par The data of Fig.\ref{fig:FigS11}(c) becomes more informative on a square root presentation of linearly (non-logarithmic) scaled $P_{rf}$ coordinate as shown in Fig.\ref{fig:FigS11}(d). Here, the LSM PR is linearly proportional to the circulating rf currents below critical value.

\end{widetext}

\bibliography{Ref_SQUID_LSM_v3}

\end{document}